\documentclass[sigconf]{acmart}

\setcopyright{acmcopyright}
\usepackage{tikz}
\usepackage{amsmath}

\usepackage{multirow}
\usepackage{url}
\usepackage{multicol}
\usepackage{subcaption}
\usepackage{graphicx} 
\usepackage{hyperref}
\usepackage{amsmath}

\usepackage{xspace}

\usepackage{tikz} 
\usetikzlibrary{arrows,automata}
\usepackage[title]{appendix}

\hypersetup{colorlinks,linkcolor=[rgb]{.75,0,0},urlcolor=[rgb]{0.75,0,0.75},citecolor=blue,breaklinks} %

\newcommand\V[1]{{\mbox{\textit{#1}}}}

\makeatletter
  \newcommand\EatSpacesHack{\@bsphack\@esphack}
\makeatother
  \renewcommand\comment[1]{\EatSpacesHack}
  \newcommand\reviewfix[1]{\EatSpacesHack}
  \newcommand\PostSubmission[1]{\EatSpacesHack}
\makeatother

\def\Snospace~{\S{}} %
\date{}

\newcommand{\broot}{\texttt{B-root}\xspace}

\setcopyright{none}
\renewcommand\footnotetextcopyrightpermission[1]{}
\begin{document}

\title{Internet Outage Detection using Passive Analysis\\
(Poster Abstract and Poster)}

%
 % \ifisanon
 % \else
  \author{Asma Enayet}
  \affiliation {
    \institution{University of Southern California}
    \department{Information Science Institute and CS Dept.}
    \city{Los Angeles, California}
    \country{USA}
  }
  \author{John Heidemann}
  \affiliation {
    \institution{University of Southern California}
    \department{Information Science Institute and CS Dept.}
    \city{Los Angeles, California}
    \country{USA}
  }
 % \fi

\maketitle
\pagestyle{plain}

\vspace*{-1.5ex}

Outages from natural disasters, political events, software or hardware issues, and human error \cite{comprehensive} place a huge cost on e-commerce (\$66k/minute at Amazon~\cite{amazon_cost}).

\reviewfix{}
While several existing systems detect Internet outages~\cite{quan2013trinocular, richter2018advancing, shah2017disco, guillot2019chocolatine, holterbach2019blink},
  these systems often too inflexible,
  fixed parameters across the whole internet with CUSUM-like change detection.
We instead propose a system using passive data, to cover both IPv4 and IPv6,
  customizing parameters for each block to optimize
  the performance of our Bayesian inference model.

Our poster describes our three contributions:
First,
  we show how customizing parameters
  allows us often to detect outages that are at both fine timescales (5 minutes)
  and fine spatial resolutions (/24 IPv4 and /48 IPv6 blocks).
Our second contribution is to show that,
  by tuning parameters different for different blocks,
  we can scale back temporal precision to cover more challenging blocks.
Finally, we show our approach extends to IPv6 and provide the first reports of IPv6 outages.

\textbf{Approach summary:}
\reviewfix{}
Our approach uses passive traffic observations from network-wide services.
In our evaluation we consider traffic from \broot's DNS service,
  but in principal we could use data from a large website (like Wikipedia, Google, or Amazon)
  or other infrastructure (like NTP).
We build a model of historical traffic from each source to the service,
  then detect interruptions 
  that violate model history,
  using with Bayesian inference to detect outages.
Some strong sources directly support detection,
  but when possible, we correlate multiple signals from the same region to corroborate results.
When sources are weak, we can aggregate data over larger time periods
  or spatial scales to gain confidence.

\comment{}
Currently we set parameters per-block based on history.
\reviewfix{}
Future work will consider seasonal and diurnal effects.

\textbf{Detecting short outages:}
Prior work either detects 11-minute outages at fine spatial scales (typically /24 blocks,~\cite{quan2013trinocular,richter2018advancing})
  or 5-minute outages, but for coarse spatial scales (entire ASes,~\cite{shah2017disco,guillot2019chocolatine}),
  or very fast reaction but with a very large amount of input data (seconds, but requiring all TCP flows~\cite{holterbach2019blink}).
In each case, prior systems only improve temporal resolution
  by increasing active traffic, passive spatial scale or input data.
Our new approach interprets passive data
  and can employ exact timestamps of observed data,
  allowing both fine spatial and temporal precision in many cases.

Although our approach can apply to many traffic sources,
  we quantify its effectiveness for both long-duration and short-duration outages,
  by evaluate \broot~\cite{broot} as a passive data source.
We test against Trinocular active outage detection~\cite{quan2013trinocular}
  and observations from RIPE Atlas (inspired by Chocolatine~\cite{guillot2019chocolatine})
  as ground truth.
Since \broot coverage is limited,
  we compare only /24 IPv4 blocks that overlap between our observations from \broot
  and Trinocular's observations.

We evaluate our accuracy in the confusion matrix in \autoref{tab:accuracy_after_correction_seconds}.
We define a false outage (\V{fo}) as a prediction of down when it’s really up in Trinocular, with analogous definitions of false availability (\V{fa}), true availability (\V{ta}), and true outages (\V{to}).
\reviewfix{}
High precision means the outages that we report are true,
  and strong recall means we correct estimate duration.
TNR suggests that we often find shorter outages than Trinocular.
This difference may be due to Trinocular's precision
  ($\pm 330$\,s),
  while using exact timestamps of data allows us to be more precise
  and often shorter.
To avoid uncertainty and show our model works for finding the maximum number of outages we also test on only dense data having high frequency of traffics in \autoref{tab:accuracy_after_correction_dense}.   
This shows that we have very good precision and recall for very dense blocks and TNR shows that we can detect 96\% of the outages.
\comment{}
\comment{yes, keeping the table is DEFINITELY stronger, except that we don't have it in events,
and in time it looks horrible.  So better to say ``in progress'' than to say a bad number (that we know we will replace) in print. ---johnh 2022-09-12}
\reviewfix{}
Evaluation of short outages is challenging:
  precision of $\pm 180$\,s hides
  differences in uncertainty
  for short outages (300\,s or less).
The poster will compare short outages
  by events (not time) to factor out imprecision in timing.

In the \autoref{tab:short_outages} we report outage events that are 5 minutes in length for both \broot and RIPE data. 
Our results show that we have great precision (0.9979) and recall (0.9479) indicating our model has good accuracy in long-duration outages (11 minutes or more). 
Similarly, we have great precision (0.9769) and recall (0.9453) for short-duration outages (5 minutes or more). 
Our measurements show that on 2019-01-10,
  around 5\% of total blocks that have 5\,minute outages that were
  not seen in prior work.
These short outages add up---when we add the outages from 5 to 11\,minutes that were previously omitted to observations,
  we see that total outage duration increases by 20\%.

%
 % \iffalse
 % \fi
%
\begin{table}
\begin{center}
\resizebox{0.45\textwidth}{!}{%
\begin{tabular}{ c|c c|c} 
  \textbf{Observation}& \multicolumn{2}{c|}{\textbf{Ground truth (Trinocular)}}&\\
 (\broot)& availability (s) & outage (s) &\\
 \hline
 availability & TP = ta = 52525765695 & FP = fa = 2471178 & Precision 0.9999\\
 outage & FN = fo = 78163261 & TN = to = 13147965 \\
 \hline
 &Recall 0.9985& TNR 0.84178\\
\end{tabular}
}
\end{center}
\caption{Confusion matrix for long-duration outages (in seconds)}
\label{tab:accuracy_after_correction_seconds}
\end{table}
\begin{table}
\begin{center}
\resizebox{0.45\textwidth}{!}{%
\begin{tabular}{ c|c c|c} 
  \textbf{Observation}& \multicolumn{2}{c|}{\textbf{Ground truth (Trinocular)}}&\\
 (\broot)& availability (s) & outage (s) &\\
 \hline
 availability & TP = ta = 7644527262 & FP = fa = 77152 & Precision 0.99\\
 outage & FN = fo = 387011 & TN = to = 2233042 \\
 \hline
 &Recall 0.99& TNR 0.96\\
\end{tabular}
}
\end{center}
\caption{Confusion matrix for long-duration outages on dense blocks (in seconds)}
\label{tab:accuracy_after_correction_dense}
\end{table}
\begin{table}
\begin{center}
\resizebox{0.45\textwidth}{!}{%
\begin{tabular}{ c|c c|c} 
  \textbf{Observation}& \multicolumn{2}{c|}{\textbf{Ground truth (RIPE)}}&\\

 (\broot)& availability (events) & outage (events) &\\
 \hline
 availability & 4445 & 105  &Precision 0.97692\\
 outage & 257 & 290 \\
 \hline
 &Recall 0.9453& TNR 0.7341\\
\end{tabular}
}
\end{center}
\caption{Confusion matrix for short-duration outages (events)}
\label{tab:short_outages}
\end{table}
\textbf{Optimizing across a diverse Internet:}
Because the Internet is so diverse, outage detection systems need
  to be tuned to operate differently for differently-behaving regions.
We describe the first passive system that optimizes parameters to each block
  to provide fine spatial and temporal precision when possible,
  but falling back on coarser temporal precision when necessary.
By contrast, prior passive systems are often homogeneous, using
  the same parameters across all block and therefore providing only
  coarse spatial coverage (at the country or AS level~\cite{guillot2019chocolatine},
  or decreasing coverage.
We exploit the ability to trade-off between spatial and temporal precision.
We customize parameters to treat each blocks differently, allowing different regions to have different temporal and spatial precision. 
As a result we can get the coverage of more sparse blocks when we employ less temporal precision.

\begin{figure}
\centering
\includegraphics[height=4.5cm]{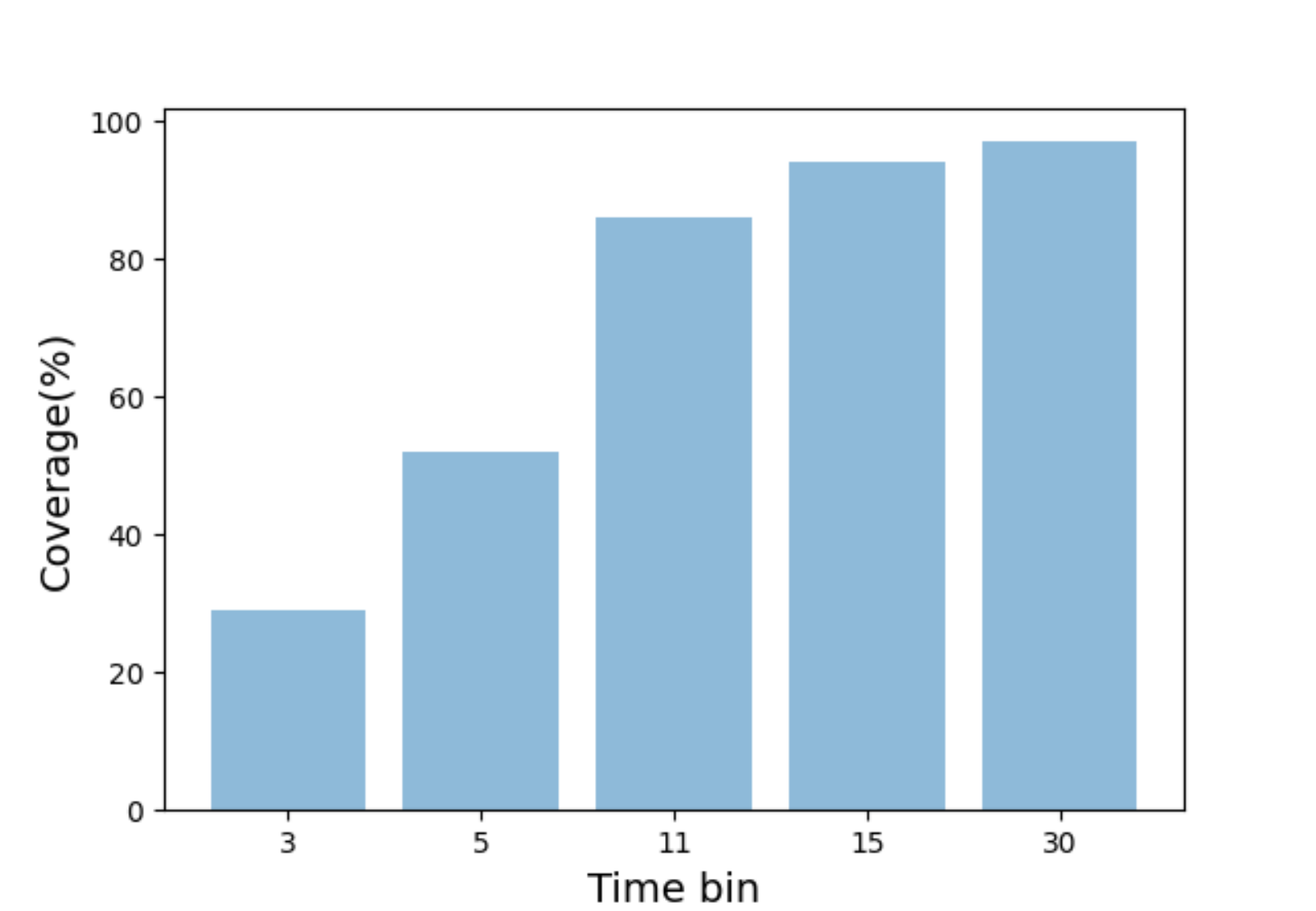}
\caption{Trading temporal and spatial precision.}
\label{fig:time_vs_coverage}
\end{figure}

In \autoref{fig:time_vs_coverage}
 we evaluate this trade-off,
 showing that we have good precision for the dense blocks and less precision for the sparse blocks.
We can observe almost 90\% of \broot blocks in one day including sparse blocks if we take longer time bin which means less temporal precision.
Therefore, the user can choose and get the best precision or coverage depending on the characteristics of the data.

\textbf{IPv6:}
IPv6 is 
  a growing part of the Internet today,
  and of course it has outages,
  but prior outage-detection systems have not extended to IPv6.
Prior active monitoring systems
  cannot possibly probe all unicast IPv6 address,
  since $2^{128}$ addresses requires centuries to scan,
  and privacy-preserving addressing makes most client addresses ephemeral.
Our new approach extends coverage to IPv6,
  by analyzing passive data,
  allowing the active addresses to come to us.

\begin{figure}
  \begin{subfigure}[b]{0.45\columnwidth} 
    \includegraphics[width=\linewidth, height =4cm]{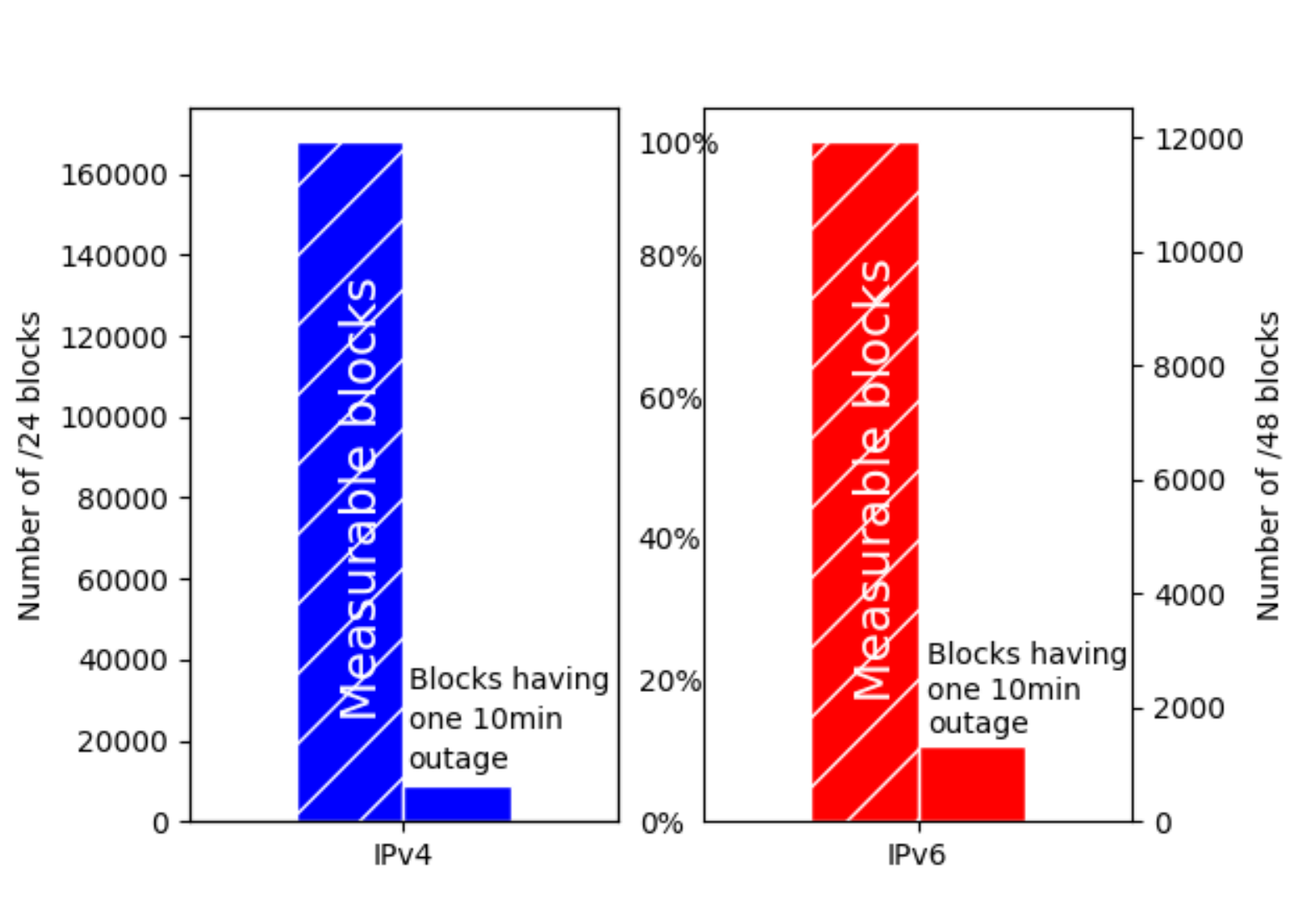}
    \caption{Outage}
    \label{fig:ipv6_outage}
  \end{subfigure}
\quad
  \begin{subfigure}[b]{0.45\columnwidth}
    \includegraphics[width=\linewidth, height =4cm]{graphs/ipv4ipv6_coverage.pdf}
    \caption{Coverage}
    \label{fig:ipv6_coverage}
     \end{subfigure}  
     \caption{IPv6 coverage is similar to IPv4}
\end{figure}

We evaluate our IPv6 coverage based on one representative day of
  passive data from \broot,
  comparing results in IPv4 and IPv6.
In \autoref{fig:ipv6_outage} we see 11,918 /48 IPv6 blocks that are measurable (they have enough data to provide a reliable outage signal),
  and we see at least one 10\,minute outages in 1338 (12\% of measurable blocks).
By comparison, the same system sees 167,851 /24 IPv4 blocks that are measurable,
  and 8689 /24 IPv4 blocks have one 10\,minute outage  (5.5\% of measurable blocks).
The absolute number of IPv4 outages is larger than IPv6 because
  there are far more measurable IPv4 blocks.
However, the outage \emph{rate} (the percentage measurable blocks with outages)
  for IPv6 seems somewhat greater than for IPv4,
  suggesting IPv6 reliability can improve.

Our coverage in IPv6 is surprisingly large:
\autoref{fig:ipv6_coverage} compares coverage relative to best prior system.
For IPv6, our approach with \broot sees about 12,765 IPv6 /48 blocks,
  17\% of the  74,373 /48 blocks in the Gasser IPv6 hitlist~\cite{gasser2018clusters}.
This coverage is similar to what we see in IPv4,
  where the 1M /24s blocks our system with \broot
  is about 20\% of the 5.1M in Trinocular.
In both cases, \broot coverage is limited (it sees only recursive resolvers),
  but it seems about the same fraction of IPv6 as IPv4.
We expect to add additional passive sources to increase IPv6 coverage.

Although our work is still in progress,
  our early results suggest a significant
  advance in the ability to observe both long and short outages of the network edge
  for both IPv4 and IPv6.
We show that users can tradeoff between spatial and temporal precision depending on the type of data.
We show our IPv6 provides coverage similar to our IPv4 coverage,
  and  provide the first report IPv6 outages.

\emph{This work was supported by NSF grant CNS-2007106 (EIEIO).}
\bibliographystyle{abbrv} 
% \bibliography{ref} 

%
\begin{appendices}
\end{appendices}
\begin{center}
    \includegraphics[page=1,width=\textwidth,height=\textheight,keepaspectratio]{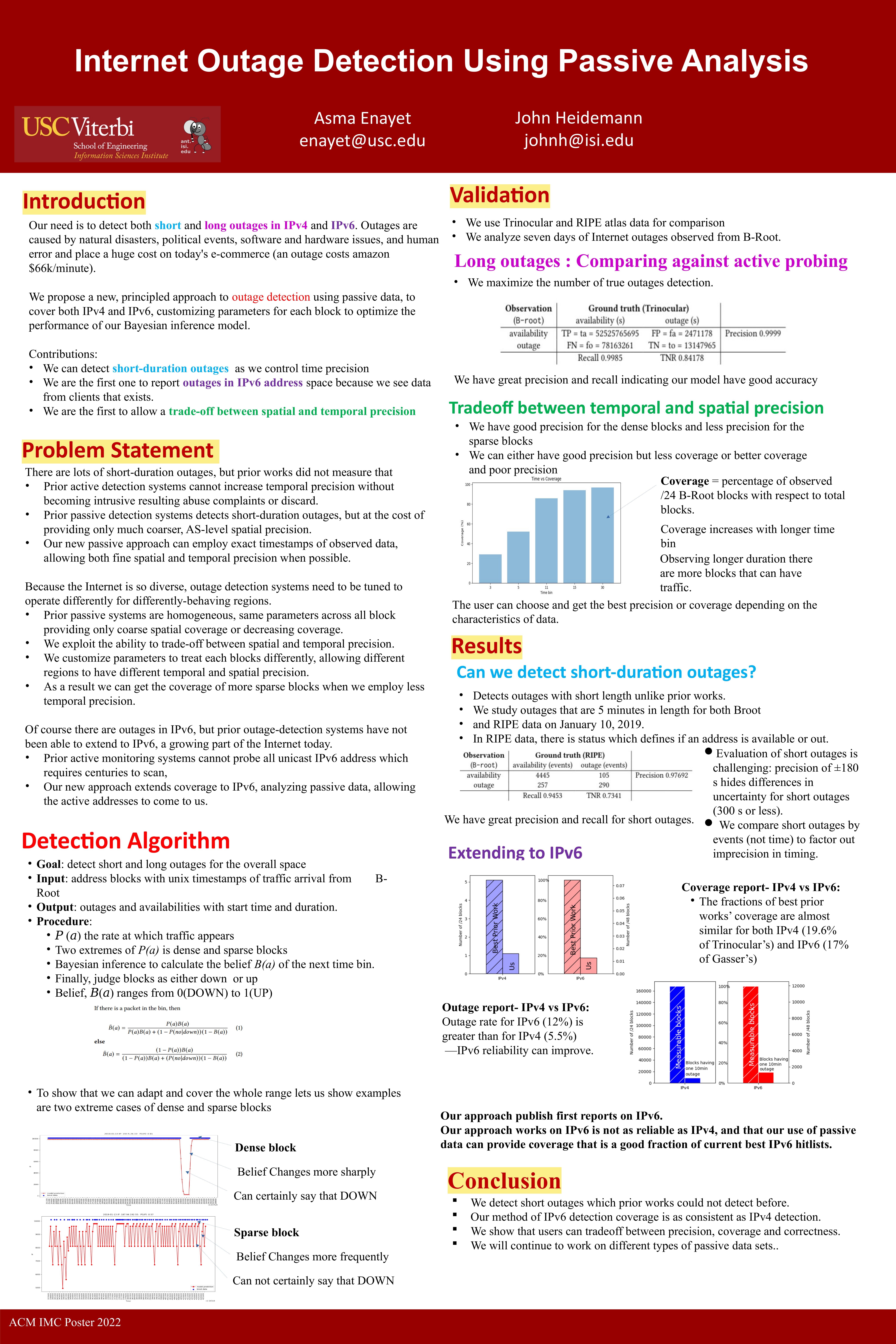}
\end{center}

\end{document}